\documentclass[12pt]{article}

\usepackage{a4p}
\usepackage{amssymb}
\usepackage{epsfig}
\usepackage{cite}
\usepackage{pennames}
\usepackage{rotating}
\usepackage{array}

\newcommand{\inmath}[1] {\ifmmode#1\else$#1$\fi}
\newcommand{\definmath}[2]{\def#1{\ifmmode#2\else$#2$\fi}}

\newcommand{\PZ}        {\mbox{$\mathrm{Z}$}}  
\definmath{\PWpm}       {\mathrm{W}^{\pm}}     
\definmath{\Plp}        {\ell^{+}}             
\definmath{\Plm}        {\ell^{-}}             
\definmath{\Plpm}       {\ell^{\pm}}           
\definmath{\Pgtp}       {\tau^{+}}             
\definmath{\Pgtm}       {\tau^{-}}             
\definmath{\Pgtpm}      {\tau^{\pm}}           
\definmath{\Pgn}        {\nu}                  
\definmath{\Pagn}       {\overline{\nu}}       
\definmath{\Pf}         {\mathrm{f}}
\definmath{\Paf}        {\overline{\mathrm{f}}}
\definmath{\Pq}         {\mathrm{q}}
\definmath{\Paq}        {\overline{\mathrm{q}}}
\definmath{\Pu}         {\mathrm{u}}
\definmath{\Pau}        {\overline{\mathrm{u}}}
\definmath{\Pd}         {\mathrm{d}}
\definmath{\Pad}        {\overline{\mathrm{d}}}
\definmath{\Ps}         {\mathrm{s}}
\definmath{\Pas}        {\overline{\mathrm{s}}}
\definmath{\Pc}         {\mathrm{c}}
\definmath{\Pac}        {\overline{\mathrm{c}}}
\definmath{\Pb}         {\mathrm{b}}
\definmath{\Pab}        {\overline{\mathrm{b}}}
\definmath{\Pt}         {\mathrm{t}}
\definmath{\Pat}        {\overline{\mathrm{t}}}
\definmath{\Pap}        {\overline{\mathrm{p}}}
\definmath{\Pan}        {\overline{\mathrm{n}}}
\definmath{\PaD}        {\overline{\mathrm{D}}}
\definmath{\PaDz}       {\overline{\mathrm{D}}^{0}}
\definmath{\PaB}        {\overline{\mathrm{B}}}
\definmath{\PaBz}       {\overline{\mathrm{B}}^{0}}

\definmath{\dedx}{{\rm d}E/{\rm d}x}

\newcommand{\qqbar}     {\Pq\Paq}
\newcommand{\qq}     {\Pq\Paq}

\newcommand{\WW}{\ensuremath{\mathrm{W}^+\mathrm{W}^-}}

\definmath{\GeV}        {\mathrm{GeV}}
\definmath{\GeVc}       {\mathrm{GeV}\!/c}
\definmath{\GeVcc}      {\mathrm{GeV}\!/c^2}
\definmath{\MeV}        {\mathrm{MeV}}
\definmath{\MeVc}       {\mathrm{MeV}\!/c}
\definmath{\MeVcc}      {\mathrm{MeV}\!/c^2}
\definmath{\keV}        {\mathrm{keV}}
\definmath{\keVcm}      {\mathrm{keV}\!/\mathrm{cm}}

\newcommand{\PhysLett}  {Phys.~Lett.}

\newcommand{\PhysRev}   {Phys.~Rev.}
\newcommand{\NPhys}     {Nucl.~Phys.}
\newcommand{\NIM}       {Nucl.~Instr.\ Meth.}
\newcommand{\ZPhys}     {Z.~Phys.}

\newcommand{\CPC}       {Comput. Phys. Commun.}

\newcommand{\OPALColl}  {OPAL Collaboration}

\newcommand{\ee}        {\mathrm{e}^+\mathrm{e}^-}

\def\gappeq{\ensuremath{\mathrel{ \rlap{\raise.5ex\hbox{>}}
                    {\lower.5ex\hbox{\sim}}}}}
\def\lappeq{\ensuremath{\mathrel{ \rlap{\raise.5ex\hbox{<}}
                    {\lower.5ex\hbox{\sim}}}}}

\def\etal               {\mbox{{\it et al.}}}

\newcommand{\zz}        {\PZ \PZ}

\newcommand{\zg}    {\PZ \gamma^*}
\newcommand{\gsgs}    {\gamma^* \gamma^*}

\newcommand{\mm}        { \mu^+ \mu^-}

\newcommand{\qqlnu}     { \mathrm{q \bar{q}} \ell \nu}

\newcommand{\qqee}      { \qqbar {\rm e}^{+} {\rm e}^{-} }
\newcommand{\qqmm}      { \qqbar \mu^{+} \mu^{-} }

\newcommand{\mee}       {m_{\rm e^+e^-}}
\newcommand{\mz}       {m_{\rm Z}}

\newcommand{\mmumu}     {m_{\mu^+\mu^-}}

\newcommand{\mqq}       {m_{\mathrm{q \bar{q}}}}

\begin{document}
\begin{titlepage}
\begin{center}
{\Large 
   EUROPEAN ORGANIZATION FOR NUCLEAR RESEARCH}

\end{center}
 \begin{center}{\large
 }\end{center}\bigskip
\begin{flushright}
  CERN-EP/2002-052 \\ July 15th, 2002 \\
\end{flushright}
\bigskip\bigskip\bigskip
\begin{center}
{
\huge\bf\boldmath  
Measurement of Neutral-Current Four-Fermion Production at LEP2 

}
\end{center}
\bigskip\bigskip
\begin{center}
{\Large
The OPAL Collaboration
}
\end{center}
\bigskip\bigskip\bigskip
\begin{center}
{\large Abstract}
\end{center}
Four-fermion final states $\qqee$ and $\qqmm$ 
from neutral-current interactions
in $\ee$ collisions are studied
in the OPAL detector
at LEP at centre-of-mass energies from 183~GeV to 209~GeV.
The data analysed correspond to a total integrated luminosity
of about 650~${\rm pb}^{-1}$ recorded from 1997 to 2000.
Corresponding to the acceptance of the OPAL detector,
a signal definition is applied requiring both leptons to
have a scattering angle satisfying $|\cos \theta |< 0.95$.
Further requirements are made on the invariant masses of the
fermion pairs.
The extracted  cross-sections for the processes
$ \ee \to \qqee$  and $ \ee \to \qqmm$ 
are consistent with the expectations from the Standard Model.

\bigskip\bigskip
\bigskip\bigskip

\end{titlepage}


\begin{center}{\Large        The OPAL Collaboration
}\end{center}\bigskip
\begin{center}{
G.\thinspace Abbiendi$^{  2}$,
C.\thinspace Ainsley$^{  5}$,
P.F.\thinspace {\AA}kesson$^{  3}$,
G.\thinspace Alexander$^{ 22}$,
J.\thinspace Allison$^{ 16}$,
P.\thinspace Amaral$^{  9}$, 
G.\thinspace Anagnostou$^{  1}$,
K.J.\thinspace Anderson$^{  9}$,
S.\thinspace Arcelli$^{  2}$,
S.\thinspace Asai$^{ 23}$
% ...
}\end{center}\bigskip
\bigskip
$^{  1}$School of Physics and Astronomy, University of Birmingham,
Birmingham B15 2TT, UK
\newline
$^{  2}$Dipartimento di Fisica dell' Universit\`a di Bologna and INFN,
I-40126 Bologna, Italy
\newline
$^{  3}$Physikalisches Institut, Universit\"at Bonn,
D-53115 Bonn, Germany
\newline
% ...
$^{  5}$Cavendish Laboratory, Cambridge CB3 0HE, UK
\newline
% ...
$^{  9}$Enrico Fermi Institute and Department of Physics,
University of Chicago, Chicago IL 60637, USA
\newline
% ...
$^{ 16}$Department of Physics, Schuster Laboratory, The University,
Manchester M13 9PL, UK
\newline
% ...
$^{ 22}$Department of Physics and Astronomy, Tel Aviv University,
Tel Aviv 69978, Israel
\newline
$^{ 23}$International Centre for Elementary Particle Physics and
Department of Physics, University of Tokyo, Tokyo 113-0033, and
Kobe University, Kobe 657-8501, Japan
\newline
% ...

\section{Introduction}
\label{sec:intro}

Four-fermion processes in $\ee$ collisions have proven to be an important tool
for studying the validity of the Standard Model and looking
for signs of physics beyond. 
At centre-of-mass energies ($\sqrt{s}$) reached at LEP2, from 183~GeV 
to 209~GeV,
pair-production of the gauge bosons of the weak interaction,
the W- and Z-bosons, has been studied extensively~\cite{bib:lepw,bib:lepz}.
The cross-sections for these two processes are characterized by a steep 
increase near threshold. 
They have been measured by restricting 
the invariant masses of pairs of fermions to the W or Z mass,
thereby selecting only some of all possible four-fermion final states.
It is therefore interesting to extend the measurement into regions where the 
invariant masses of the fermion-pairs are not as restricted.
Results on this have been reported from the LEP experiments 
for centre-of-mass energies up to 183~GeV~\cite{bib:lep4f}.
The present study provides results at higher energies with greatly increased
luminosity.
This extension provides
 an especially important test for the production of new particles
with masses well below or above the W or Z mass.

\begin{figure}
\centerline{\epsfig{file=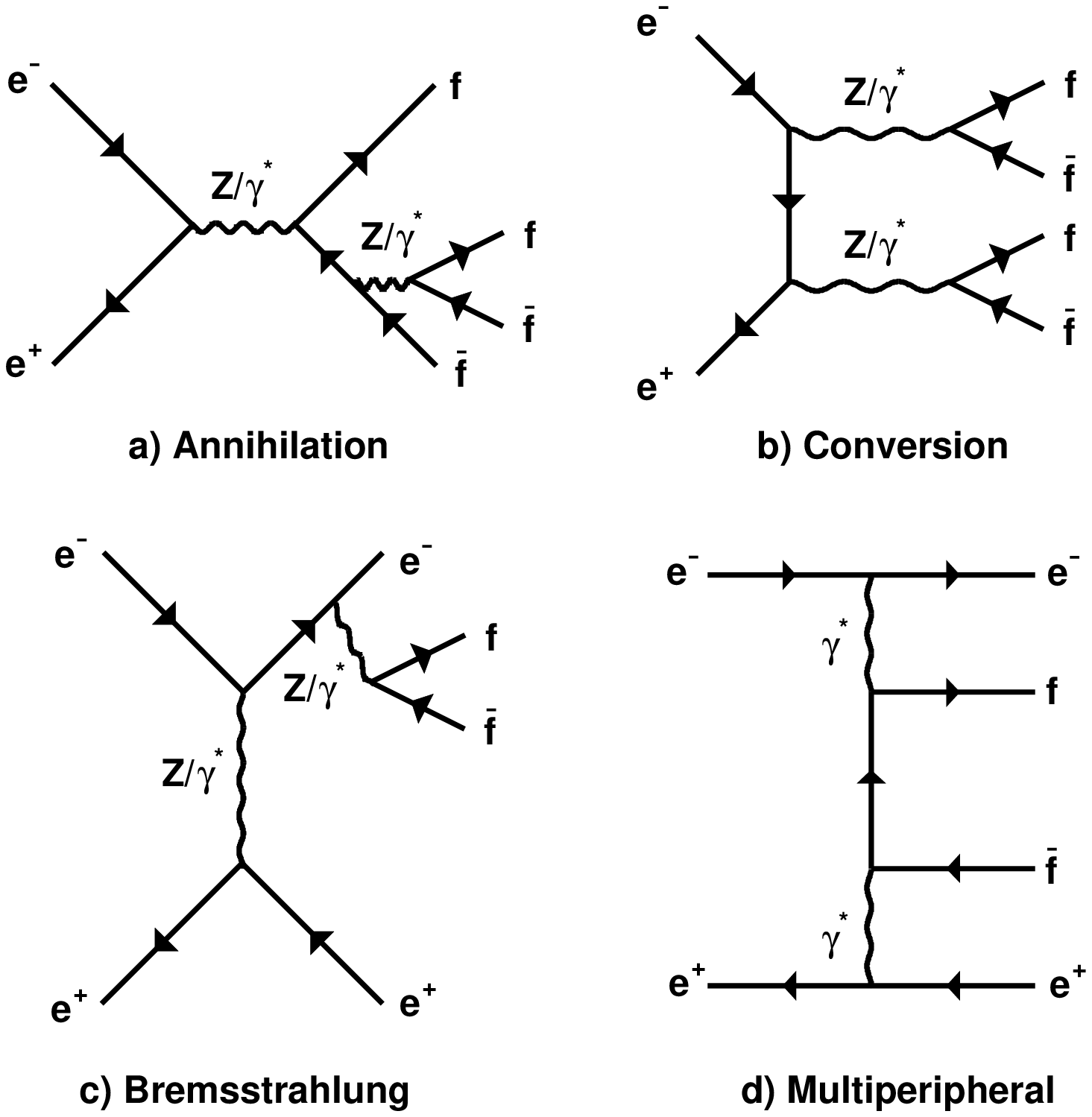,width=0.95\textwidth}} 
\caption{ Feynman diagrams for four-fermion final states involving 
neutral gauge-boson exchange: a) annihilation, b) conversion, 
c) bremsstrahlung and d) multi-peripheral production.
a) and b) involve $s$-channel 
diagrams, c) is a $t$-channel diagram for  signal.
Events stemming from d) are regarded as background.
}
\label{fig:plots0}
\end{figure}

In this paper we describe a selection for $\qqee$ and $\qqmm$ 
final states that
is sensitive to all invariant masses of fermion pairs.
Restrictions are made only for low hadronic ($\qq$) mass values 
to avoid the region
of hadronic resonances and at very small masses of the 
electron\footnote{Charge conjugation is implied throughout this paper.}
pair ($\ee$), where low efficiencies compromise the measurement.
The final states of interest are produced via  $s$-channel
annihilation of the two incoming 
electrons into a Z or $\gamma^*$ and the radiation of a Z or $\gamma^*$
from either an incoming electron or an outgoing fermion, for example as shown 
in Figures~\ref{fig:plots0}a) and b).
For the case of $\qqee$ final states, there is also a $t$-channel 
contribution, as shown in Figure~\ref{fig:plots0}c).
In a large fraction of the $t$-channel events,
one or both electrons cannot be detected, as their
scattering angle is small and outside of the acceptance of the detector.
We therefore define the following kinematic bounds to classify events as 
signal:
\begin{itemize} 
\item
The polar angles $\theta$, defined with respect to the incoming electron beam,
of both leptons have to satisfy $|\cos \theta |< 0.95$.
\item 
The invariant mass of the $\qq$ system has to be larger than 
5~GeV in order to avoid the complex region of
hadronic resonances.
\item 
For the $\qqee$ final state,
the invariant mass of the $\ee$ system has to be larger than 
2~GeV. For invariant masses below 2~GeV, the efficiency is too
small  to provide a reliable measurement.
There is no restriction imposed on the invariant mass of the $\mm$ system 
in the $\qqmm$ analysis.
\item
Events stemming from multi-peripheral diagrams,
involving the exchange of two photons (Figure~\ref{fig:plots0}d)),
are not considered as signal. 
The interference between the signal diagrams and the multi-peripheral diagrams
is small compared to the signal cross-section 
and is neglected.
\end{itemize}

In about 650~${\rm pb}^{-1}$ of data recorded with the
OPAL detector in the years 1997 to 2000 at centre-of-mass energies
between 183~GeV and 209~GeV, the cross-sections for the processes
$\ee \to \qqee$ and $\ee \to \qqmm$ are measured within the 
kinematic bounds described above.

The paper is organized as  follows.
First, a description of the  OPAL detector and the data sample is given.
Then, the Monte Carlo generators used to simulate
the signal and background events are described.
Following this, the analyses used to select the signal events are detailed.
After a description of the systematic studies, the 
measured cross-sections are given, together with a discussion of the results.

\section{Description of Data and Detector}
\label{simulation}

Data recorded with the OPAL detector in the years 1997 to 2000
at centre-of-mass energies between 183~GeV and 209~GeV are used in this 
analysis. The data taken in each year are analysed separately, except for
1999, where data were taken at centre-of-mass 
energies between 192~GeV and 202~GeV. Due to the large range of centre-of-mass
energies these data are subdivided into two samples, one below and one 
above 197~GeV.

A detailed description of the OPAL detector can be found 
elsewhere~\cite{detector} and only a brief description is given here.
The central detector consists of a system of tracking chambers
which provide charged particle tracking over 96\% of the full solid angle,
within a uniform magnetic field of 0.435~T parallel to the beam axis. 
The central detector is composed of a two-layer
silicon microstrip vertex detector and three different drift chambers:
a high-precision vertex chamber,
a large-volume jet chamber, and a set of $z$ chambers which provide
the track coordinates along the beam direction. 
A lead-glass electromagnetic 
calorimeter (ECAL), located outside the magnet coil,
covers the full azimuthal range with excellent hermeticity
in the polar angle range of $|\cos \theta |<0.82$ for the barrel
region, and $0.82<|\cos \theta |<0.984$ for the endcap region.
The magnet return yoke is instrumented with hadron calorimetry (HCAL),
and consists of barrel and endcap sections, along with pole-tip detectors, 
which together cover the region $|\cos \theta |<0.99$.
Four layers of muon chambers 
cover the outside of the hadron calorimeter. 
Electromagnetic calorimeters close to the beam axis
complete the geometrical acceptance down to 24 mrad, except
for regions where a tungsten shield 
designed to protect the detectors
from synchrotron radiation is located.
These calorimeters include 
the forward detectors (FD), which are
lead-scintillator sandwich calorimeters, and, at smaller angles,
silicon-tungsten calorimeters~\cite{ref:SW}
located on both sides of the interaction point.
The silicon-tungsten calorimeters are used to evaluate the luminosity 
by observing small-angle Bhabha events.

\section{Monte Carlo simulation}

Four-fermion final states for the signal processes are generated with the 
grc4f~\cite{bib:grc4f} Monte Carlo (MC) program.
For the $\qqee$ final states, both $s$-channel and $t$-channel diagrams 
contribute.
In the $t$-channel processes, the momentum 
transfer squared between the two electrons is in general small.
The relevant value of the electromagnetic coupling constant
$\alpha$ is therefore $\alpha(0) \approx 1/137$. 
For the $s$-channel diagrams, a scale of the order of $\sqrt{s}$ is important,
leading to a larger value of $\alpha(\sqrt{s}) \approx 1/128$.
To account for interference effects, the $s$- and $t$-channel diagrams
have to be generated simultaneously.
Within the given kinematic limits of the signal definition, 
the $s$-channel contribution is only 
slightly larger than that from the $t$-channel and the interference between 
the two channels is negative and of the order of 15\%. 
In grc4f v2.1, a fixed value of $\alpha$ is used in the generation of events.
To investigate the impact of the chosen value of $\alpha$  
on the selection efficiencies, two sets of 
signal Monte Carlo for $\qqee$ final states
are generated using different values, $\alpha = 1/137$ and  $\alpha = 1/128$,
leading to an average 15\% difference in the cross-section.

For $\qqmm$ events, only $s$-channel diagrams
contribute, and a value of $\alpha = 1/128$ is used.
For comparison to grc4f, EXCALIBUR~\cite{bib:excalibur} and 
KORALW~\cite{bib:koralw} are also used to produce $\qqmm$
signal events.

Events from  multi-peripheral diagrams 
with at least one electron inside the detector acceptance
are produced using the TWOGEN~\cite{bib:twogen} generator.
As a check the generator PHOJET~\cite{bib:phojet} is also used 
for multi-peripheral diagrams 
with both electrons inside the detector.
Hadronization is performed with JETSET~\cite{bib:jetset}. 
Four-fermion final states not included in the signal definition and not
stemming from  multi-peripheral diagrams are generated with grc4f.
Processes involving two fermions in the final state are simulated using
KK2f~\cite{bib:kk2f} for multi-hadronic ($\qq$) events.
PYTHIA~\cite{bib:jetset} is used as a cross-check.
For the process $\ee\to\ee$, BHWIDE~\cite{bib:bhwide} is used.
Background contributions
from other processes are found to be negligible. 
All Monte Carlo events are passed through the
full simulation of the OPAL detector~\cite{bib:gopal}, and then
subjected to the same reconstruction and analysis procedures as data.

\def\pbl{\phantom{(}} \def\pbr{\phantom{)}}

\section{Event Selection}
\label{sec:selection}

The event selection is done separately for 
$\qqee$ and $\qqmm$ final states,
but the two sets of selections are very similar.
Final states are selected according to the signal topology, 
and cuts corresponding to the signal definition given in 
Section~\ref{sec:intro} are applied.

\subsection{Selection of \mbox{ \boldmath $\qqee$} events}
\label{sec:selectionee}

The selection makes use of the signal topology of two isolated
electrons and two jets, which together sum to the total centre-of-mass energy.
A kinematic fit is performed, making use of the four constraints coming from
energy and momentum conservation.
The number of events remaining after each selection cut for data,
signal, and background, are listed in Table~\ref{tab:qqee1}.

\begin{figure}
\centerline{\epsfig{file=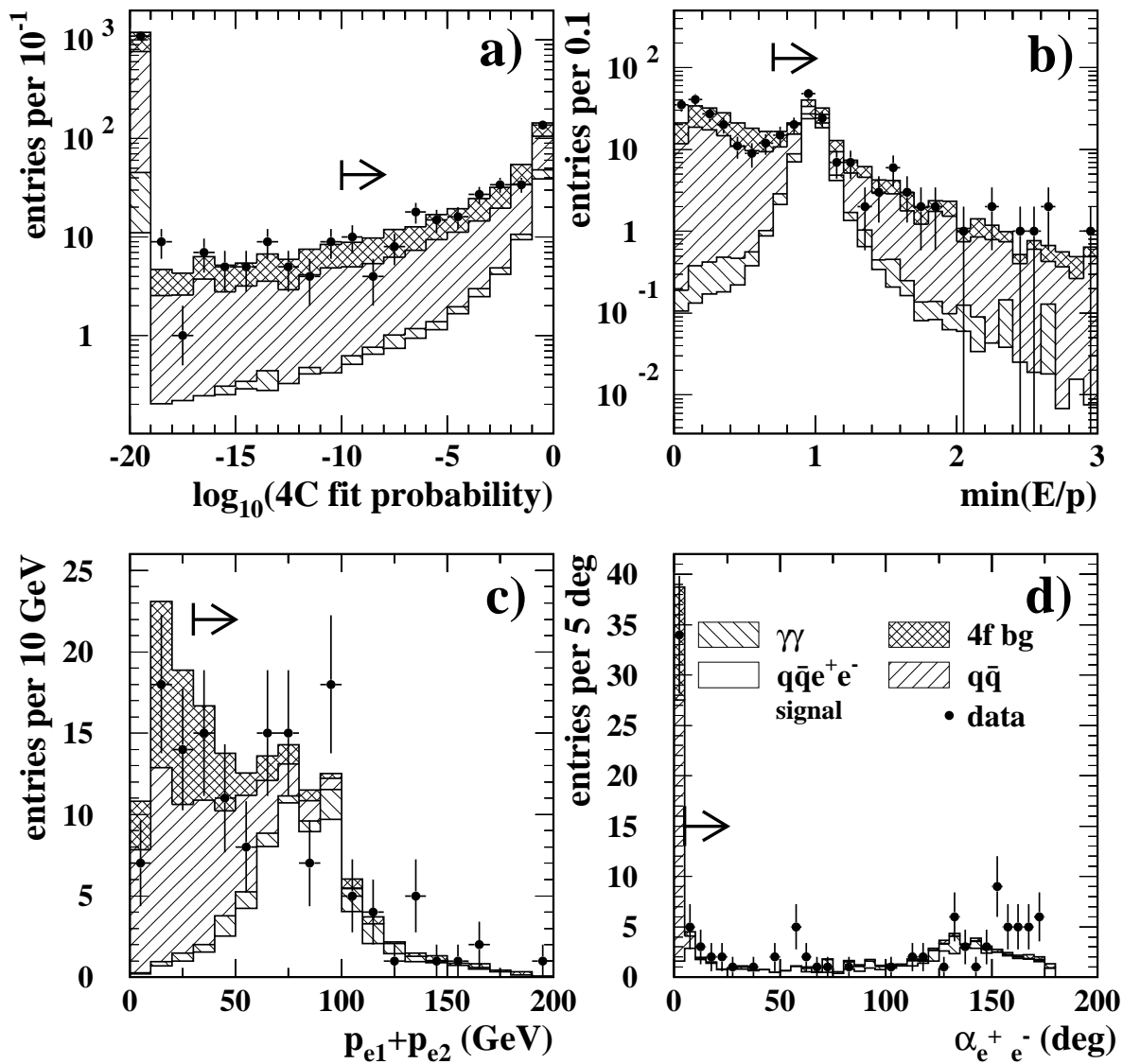,width=1.0\textwidth}}
\caption{ Distribution of
a) the logarithm of the 4C fit probability, 
b) the minimum of $E/p$ of the two electrons, 
c) the sum of the momenta of the two electrons,
d) the angle between the two electron tracks,
for data and Monte Carlo
between 183~GeV and 209~GeV for the $\qqee$ selection. The cuts have been applied
successively in a) -- d). 
Figure a) has all events remaining after the 
preselection.
The arrows point into the direction accepted by the cuts.  
The contributions from
multi-peripheral {\rm ($\gamma \gamma$)}, multi-hadronic {\rm ($\qq $)}
and four-fermion {\rm (4f)} backgrounds are shown separately.
}
\label{fig:eplot1}
\end{figure}

\begin{itemize}
\item{Cut 0:} Preselection\\ 
In the preselection at least seven tracks are required in the event
and  the visible energy, calculated using a method~\cite{bib:mtpack} 
that avoids 
double-counting of track momenta and  energy deposition in the calorimeter, 
is required to be greater than half of the centre-of-mass energy. 
At least two tracks of opposite charge, each satisfying 
the following criteria, must be present and are 
considered as  electron candidates:

\begin{itemize}
\item The absolute value of the momentum $p$ has to be greater than 2~GeV.
\item 
The electron must not be identified as arising from a 
photon conversion, i.e. the output
of the conversion neural network as described in~\cite{bib:idncon} has to be less than 0.8.
\end{itemize}

Using a neural network electron finder~\cite{ref:elecbarrel, ref:elecendcap}
an output value is calculated for each electron candidate.
From those candidates with no more than 
one track with opposite sign and momentum greater than
2~GeV within a cone of 10$^{\circ}$ half opening angle,
the candidate with the highest output value is selected
as the first electron.

From the candidates with charge opposite to that of the first electron
and no track, except for the first electron, 
within a cone of 10$^{\circ}$ half opening angle,
the one with the highest output value is taken as the second electron.
No requirement is made on the minimum output value for the electrons.

\item{Cut 1:} 4C kinematic fit\\ 
Excluding the two electron candidates selected in the preselection,
and their associated calorimeter clusters, 
the rest of the event is forced into two jets using the 
Durham~\cite{bib:durham}
jet finder.  A four-constraint kinematic fit (4C fit) is applied to the
energy and momenta of the two electron candidates and the two jets. We
use the ECAL energy and the track angles for the electron candidates
and the jet momenta as input to this
fit. The fit probability is required to be greater than $10^{-10}$
(see Figure~\ref{fig:eplot1}a). This requirement
greatly reduces background with
missing energy, for example from $\WW \to \qqlnu$.

\item{Cut 2:} Electron identification\\ 
The two electrons selected in the preselection 
are required to satisfy $E/p >0.7$,
where $E$ is the energy deposited in the electromagnetic
calorimeter associated with the track (see Figure~\ref{fig:eplot1}b).

\item{Cut 3:} Momentum cuts\\ 
The sum of the momenta of the two
electrons has to be greater than 30~GeV (see Figure~\ref{fig:eplot1}c).

\item{Cut 4:} Isolation \\
The angle between the two electron tracks is required to be 
greater than 5$^\circ$ (see Figure~\ref{fig:eplot1}d), and
the two electron tracks must not point to 
the same ECAL cluster, as otherwise their invariant
mass is difficult to reconstruct.

\item{Cut 5:} Invariant  masses of electron and quark pairs\\
Corresponding to the signal definition, 
the invariant mass of the electron pair ($\mee$)
must be greater than 2 GeV, and
the invariant mass of the quark pair ($\mqq$)
greater than 5 GeV.  
The invariant masses are obtained from the kinematic fit.

\item{Cut 6:} Multi-peripheral background \\
Multi-peripheral events typically have both electrons scattered at small
angles. To reject these events, at least one of the electrons must
have cos($\rm \theta_e)<$ 0.7, where $\rm \theta_e$ is the 
scattering angle of the electron with respect to its incoming direction.
\end{itemize}

\newcommand{\raiseb}{\raisebox{1.5ex}[-1.5ex]}
\def\pz{\phantom{0}}

\begin{table}[htb]
\begin{center}
\begin{tabular}{|l|r|r||r|r|r|r|} \hline
\multicolumn{1}{|l|}{Cut} & 
\multicolumn{1}{r|}{Data}  & 
\multicolumn{1}{c||}{Total MC} & 
\multicolumn{1}{c|}{$\qqee$} & 
\multicolumn{1}{c|}{multi-} & 
\multicolumn{1}{c|}{multi-}  &    
\multicolumn{1}{c|}{four fermion}  \\
\multicolumn{1}{|l|}{ } & 
\multicolumn{1}{r|}{ }  & 
\multicolumn{1}{c||}{ } & 
\multicolumn{1}{c|}{ } & 
\multicolumn{1}{c|}{peripheral } & 
\multicolumn{1}{c|}{hadronic }  &    
\multicolumn{1}{c|}{background }  \\
\hline

Cut 0&1450 & 1377.1 $\pm$ 7.9 & 74.6 $\pm$ 0.3 & 51.4 $\pm$ 2.3 &889.8 $\pm$ 6.5 &361.3 $\pm$ 3.8 \\
Cut 1& 303 & \pz 336.0 $\pm$ 3.4 & 60.8 $\pm$ 0.3 & 12.6 $\pm$ 0.9 & 146.0 $\pm$ 2.6 & 116.7 $\pm$ 2.0 \\
Cut 2& 148 & \pz 164.9 $\pm$ 2.3 & 58.8 $\pm$ 0.3 & 11.1 $\pm$ 0.8 & 57.1 $\pm$ 1.7 & 38.0 $\pm$ 1.3 \\
Cut 3& 109 & \pz 111.5 $\pm$ 1.8 & 56.9 $\pm$ 0.3 & 10.3 $\pm$ 0.8 & 28.4 $\pm$ 1.2 & 15.8 $\pm$ 1.1 \\
Cut 4& 75 & \pz 71.1 $\pm$ 1.2 & 56.0 $\pm$ 0.3 & 10.3 $\pm$ 0.8 & 0.6 $\pm$ 0.2 & 4.2 $\pm$ 0.9 \\ 
Cut 5& 69 & \pz 68.8 $\pm$ 1.2 & 55.6 $\pm$ 0.3 & 10.2 $\pm$ 0.8 & 0.6 $\pm$ 0.2 & 2.4 $\pm$ 0.9 \\ 
Cut 6& 58 & \pz 58.6 $\pm$ 1.0 & 51.2 $\pm$ 0.3 & 5.1 $\pm$ 0.5 & 0.6 $\pm$ 0.2 & 1.8 $\pm$ 0.8 \\ 
\hline 
\end{tabular}
\end{center}
\caption{ \it The number of events passing each successive cut for 
the $\qqee$ channel for
data between 183~GeV and 209~GeV.  
The number of events expected from Monte Carlo simulation, normalised to the 
data integrated luminosity, are also given.
The signal is simulated using $\alpha = 1/128$.
In the multi-peripheral background, only half the number of events from
TWOGEN are used.
The errors are statistical only.
}
\label{tab:qqee1}
\end{table}

After these cuts, a total of 58 events is observed in the data, with an
expectation of 58.6 events from Monte Carlo simulation.
The selection efficiency is greater than 40\% for \mbox{$\mee < \mz$},
and around 20\% for higher masses. The difference in efficiency is due to the
$s$- and $t$-channel contributions. For $\mee > \mz$, the $t$-channel is 
dominant.
Here, the scattered electrons are forward peaked, and therefore have 
a lower efficiency than in the $s$-channel.

\subsection{Selection of \mbox{ \boldmath $\qqmm$}  events}
\label{sec:selectionmm}

\begin{figure}
\centerline{\epsfig{file=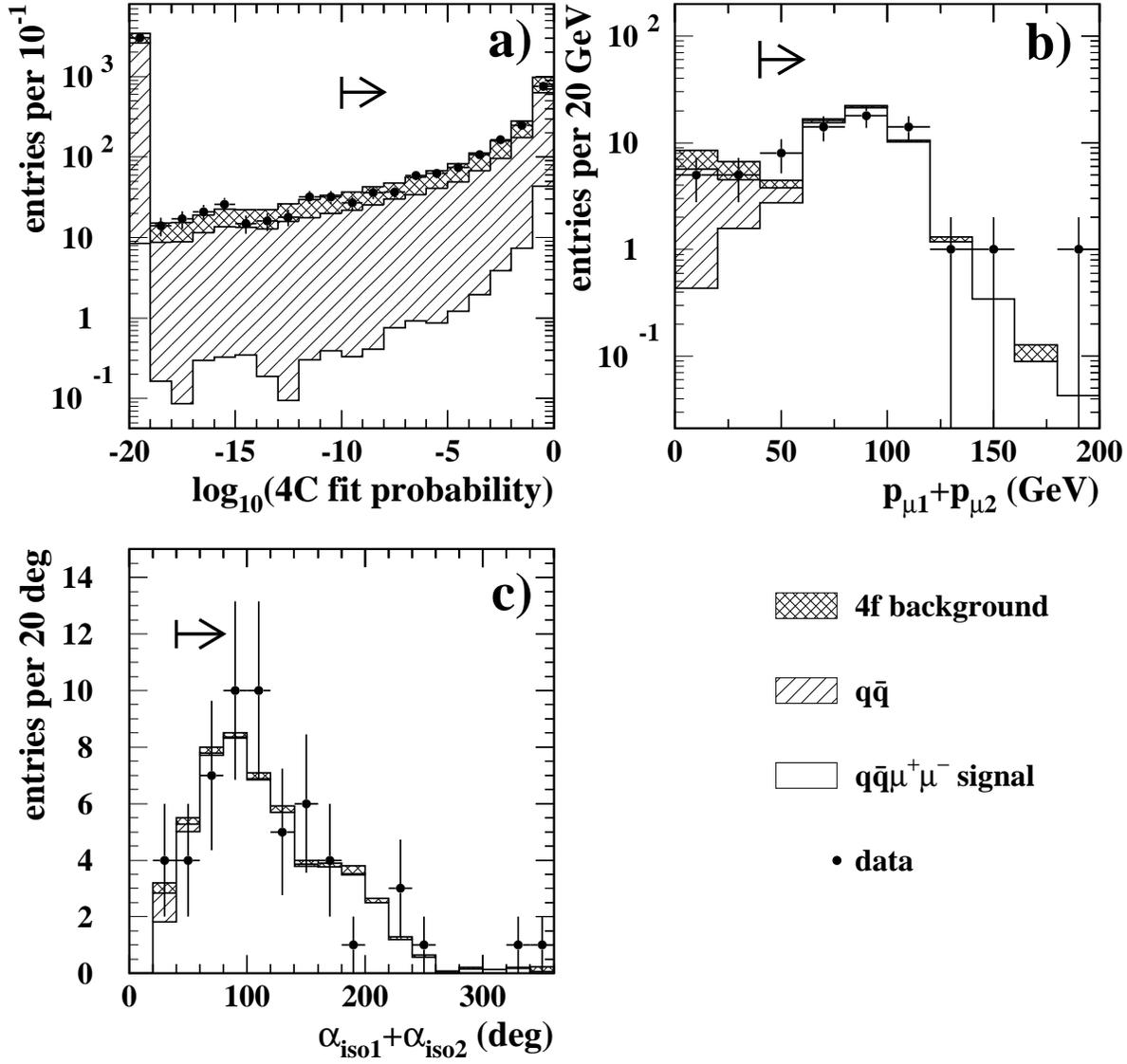,width=1.0\textwidth}}
\caption{ Distribution of
a) the logarithm of the 4C fit probability, 
b) the sum of the momenta of  the two muons, and 
c) the sum of the isolation angles of the two muons
for data and Monte Carlo
between 183~GeV and 209~GeV for the $\qqmm$ selection. 
The cuts have been applied successively in a) -- c). 
Figure a) has all events remaining after the 
preselection.
The arrows point into the direction accepted by the cuts.  
The contributions from multi-hadronic {\rm ($\qq $)}
and four-fermion {\rm (4f)} backgrounds are shown separately.
}
\label{fig:plots1}
\end{figure}

This selection is similar to the one applied for the $\qqee$ final states, 
making use of the signal topology where two muons replace the two electrons. 
The number of events, after each cut for data,
signal and background are listed in Table~\ref{tab:qqmm1}.
\begin{itemize}
\item{Cut 0:} Preselection\\ 
In the preselection, a  multiplicity of at least seven 
tracks is required.
The visible energy has to be greater than half of the centre-of-mass energy. 

From all tracks with momentum greater than 5~GeV and
no more than 
one track with opposite sign and momentum greater than
2~GeV within a cone of 10$^{\circ}$ half opening angle,
the one with the highest momentum is selected as the first muon candidate. 

From the tracks with charge opposite to that of the first muon candidate
and no track, except for the first muon candidate, 
within a cone of 10$^{\circ}$ half opening angle,
the one with the highest momentum is taken as the second muon candidate.

\item{Cut 1:} 4C kinematic fit\\ 
Excluding the two muon candidates selected in the preselection,
and their associated calorimeter clusters, the rest of the event is
forced into two jets using the Durham~\cite{bib:durham} jet
finder.  A four-constraint kinematic fit (4C fit) is applied to the
energy and momenta of the two muon candidates and the two jets. 
The track momenta of the muon candidates
and jet momenta are used as input to this
fit. The fit probability is required to be greater than $10^{-10}$
(see Figure~\ref{fig:plots1}a). 

\item{Cut 2:} Muon identification \\
A muon identification criterion is applied to the two selected 
muon candidates.
Muon identification makes use of three methods:
\begin{itemize}
\item 
Tracks are considered as muon candidates if their trajectories 
match to a track segment
in the muon chambers~\cite{bib:muonid1}, \cite{bib:muonid11}. 
\item Muons can also be identified by a selection that uses  
information from the HCAL and ECAL 
clusters~\cite{bib:muonid2},\cite{bib:muonid3}.
\item Tracks associated with an ECAL cluster with energy smaller than 2 GeV 
are selected as muon candidates.  
\end{itemize}

The first muon can be accepted using any of the above three selections.
The second muon is accepted if it fulfils the first condition,
or if it fulfils either of the other two conditions and the two following 
isolation criteria: the angle between the two muons must be greater than
 $10^{\circ}$, and the angle between the second muon and any other track, 
except for the first muon, must be greater than $30^{\circ}$.

\item{Cut 3:} Momentum cuts\\ The sum of the momenta of the two
muons has to be greater than 40~GeV (see
Figure~\ref{fig:plots1}b).

\item{Cut 4:} Isolation angles\\
The sum of the isolation angles of the two muons has to be 
greater than 40$^\circ$ (see Figure~\ref{fig:plots1}c).  
The isolation angle is the angle between the muon and any other track 
with the exception of the other muon.

\item{Cut 5:} Invariant mass of quark pair \\
Corresponding to the signal definition, 
the invariant mass of the quark pair must be greater than 5 GeV.  
\end{itemize}

\begin{table}[htb]
\begin{center}
\begin{tabular}{|l|r|r||r|r|r|} \hline
\multicolumn{1}{|l|}{Cut} & 
\multicolumn{1}{r|}{Data}  & 
\multicolumn{1}{c||}{Total MC} & 
\multicolumn{1}{c|}{$\mathrm{ q\bar{q}}\mu^+\mu^-$  } & 
\multicolumn{1}{c|}{multi-hadronic} &    
\multicolumn{1}{c|}{four-fermion} \\
& & & & &
\multicolumn{1}{c|}{background}  \\
\hline
Cut 0&4809 &4789.2 $\pm$ 14.9 & 71.9 $\pm$ 1.4 & 2972.3 $\pm$ 12.0 &1745.0 $\pm$ 8.6 \\
Cut 1& 1575 & \pz 1598.7 $\pm$ 8.0 & 61.3 $\pm$ 1.3 & 834.7 $\pm$ 6.4 & 702.7 $\pm$ 4.7 \\
Cut 2& 67 &\pz 70.1 $\pm$ 1.4 & 53.3 $\pm$ 1.2 & 7.9 $\pm$ 0.6 & 8.9 $\pm$ 0.5 \\
Cut 3& 57 & \pz 55.8 $\pm$ 1.3 & 51.3 $\pm$ 1.2 & 1.5 $\pm$ 0.3 & 3.0 $\pm$ 0.3\\
Cut 4& 53 & \pz 52.6 $\pm$ 1.3 & 49.4 $\pm$ 1.2 & 0.5 $\pm$ 0.2 & 2.7 $\pm$ 0.3 \\
Cut 5& 52 & \pz 52.3 $\pm$ 1.3 & 49.4 $\pm$ 1.2 & 0.5 $\pm$ 0.2 & 2.4 $\pm$ 0.3 \\
\hline
\end{tabular}
\end{center}
\caption{\it The number of events passing each successive cut for 
the $\qqmm$ channel for data between 183~GeV and 209~GeV.  
The number of events expected from Monte Carlo simulation, normalised to the 
data integrated luminosity, are also given.
The errors are statistical only.
}
\label{tab:qqmm1}
\end{table}

After these cuts a total of 52 events is observed in the data and
52.3 events are expected from Monte Carlo simulation, mainly originating 
from signal, with only little contribution from background.
The selection efficiency ranges from 30\% for muon invariant masses ($\mmumu$)
below 2~GeV, and is above 60\% for masses above 30~GeV.

\section{Systematic Uncertainties}
\label{sec:systematic}

Systematic uncertainties result from the determination of signal efficiencies 
and background levels, both of which are estimated from Monte Carlo samples.
For both types of
samples the agreement between the simulation and the data was
investigated, and the difference taken into account as a systematic
uncertainty.
In addition, for the signal efficiency the predictions from the different
Monte Carlo generators, and the dependence on the input parameters,
were checked, and any differences in selection efficiencies taken into
account as a systematic error.

\subsection{Selection efficiency}

Systematic uncertainties on the efficiency are calculated by varying
the cuts used in the selection.
Any difference in efficiency for the altered cut is taken
as a systematic error.
Below follows a more detailed description of how each cut was 
varied.

\begin{itemize}
\item The preselection requires a high multiplicity. The impact of a change of 
$\pm 1$ track in the preselection was studied for the signal MC.

\item The isolation angle for the leptons is required to be greater than 10$^
\circ$. The uncertainty in the measurement of the angles in the jet chamber 
is  about 0.1$^\circ$.
This has to be  multiplied by $\sqrt{2}$ for the 
measurement of the angle between two tracks.
The preselection cut is varied by this value 
to gauge its impact on the result.

\item 
A direct shift of 0.2$^\circ$ 
is applied to the angle between the two 
electrons for the  $\qqee$ selection 
to account
for possible biases in the angular reconstruction. This value was determined
from studies comparing the  angular reconstruction in the tracking and 
calorimeters.

\item For the sum of the lepton momenta, the $E/p$ value in case of the 
electrons,
and the sum of the isolation angles of the muons, the distribution of the 
variable is compared between data and MC. The MC distribution is
corrected to the data and the change in efficiency is taken as
a contribution to the systematic error. 

\end{itemize}

The systematic errors are calculated at each centre-of-mass energy.
No energy dependence is observed, therefore values derived from a comparison
of the combined data and Monte Carlo samples are used for all energies.
These values are shown in Table~\ref{tab:syst}.

\begin{table}[htp]
\begin{center}
\begin{tabular}{|l||c|r|r|r||r|r|r|} \hline
& \multicolumn{4}{|c||}{$\qqee$}& \multicolumn{3}{c|}{$\qqmm$} \\
\hline
& signal & \multicolumn{1}{c|}{multi-} & \multicolumn{1}{c|}{multi-} & 
four-fermion & 
signal &\multicolumn{1}{c|}{multi-} & four-fermion  \\
& eff.  & peripheral & hadronic & \multicolumn{1}{c||}{background} &  
\multicolumn{1}{c|}{eff.}  & hadronic & \multicolumn{1}{c|}{background} \\ 
\hline
$\sf N_{tr}$& 
2.1 \% & 4.6 $\%$ & 0 $\%$ & 0 $\%$ &
0.8 \% & 0 $\%$ & 10.6 $\%$ \\ 
\hline  
$\rm \alpha_{iso}$ & 
0.3 $\%$ &2.3 $\%$ & 0 $\%$ & 0.3 $\%$ &
0.2 \%& 11.1 $\%$ & 0.6 $\%$ \\ 
\hline 
$\rm \alpha_{ee}$ & 
0.2 $\%$  & 0 $\%$ & 0 $\%$ & 0 $\%$ &
-- & -- & -- \\ 
\hline 
$E/p$ & 
1.2 $\%$ & 1.5 $\%$ & 18.8 $\%$ & 4.7 $\%$ &
-- & -- & -- \\ 
\hline 
$\rm \sum {p_{\ell}}$ & 
0.1 $\%$ & 0  $\%$ & 6.3 $\%$ & 2.8 $\%$ &
0.8 \% & 22.2 $\%$ & 3.8 $\%$\\ 
\hline 
$\rm \sum {\alpha_{iso}}$ &
-- & -- & -- & -- &
1.3 \%  & 5.6 $\%$ & 1.9 $\%$\\ 
\hline 
\hline 
total error & 
2.4 $\%$ & 5.3 $\%$ & 19.8 $\%$ & 5.5 $\%$ &
1.7 \% & 25.5 $\%$ & 11.5 $\%$ \\ 
\hline
\hline
\multicolumn{1}{|c||}{MC} & & & & & & & \\
generators & 
10.7 \% & 100 \% & 4.5 \% & -- &
7.8 \% & 9.6 \% & -- \\
\hline
\end{tabular}
\end{center}
\caption{\it Relative systematic errors for the signal efficiency
and background Monte Carlo samples from the
systematic uncertainty studies on the number of tracks $\sf N_{tr}$,
the isolation angle of the leptons $\rm \alpha_{iso}$, 
the opening angle between the two electrons $\rm \alpha_{ee}$,
the $E/p$ value,
the sum of the lepton momenta $ \sum \rm {p_{\ell}}$, 
and the sum of the muon isolation angles $\rm \sum {\alpha_{iso}}$.
The total of these uncertainties is listed as the total error.
In the last line the uncertainty from the comparison of
Monte Carlo generators is given.
}
\label{tab:syst}
\end{table}

For the $\qqee$ selection, 
the difference in efficiency from the Monte Carlo samples generated with
values of $\alpha = 1/128$ and $\alpha = 1/137$ is 5.2\%.
For the calculation of the cross-section, the average efficiency
of the two samples is used, and assigned a systematic error of 2.6\%.
To determine the systematic uncertainty arising from use of the 
grc4f generator, a comparison with EXCALIBUR is done. 
In EXCALIBUR, events can only be generated for four-fermion final states
including  multi-peripheral diagrams. 
For this reason, additional events are generated
with grc4f, also including  multi-peripheral diagrams. The selection 
efficiency for these two event samples is compared. A difference 
of 10.7\% is found and is taken as a systematic error.
The main reason for this difference is a larger fraction of events
with small lepton scattering angles  in EXCALIBUR than in grc4f.
This type of event has a much smaller selection efficiency than events with 
large scattering angles and consequently leads to the observed difference.

For the $\qqmm$ selection, 
the expected cross-sections 
at each centre-of-mass energy agree well for the three 
MC generators grc4f, EXCALIBUR and KORALW.
There is, however, a difference observed for the selection efficiencies.
While the EXCALIBUR and KORALW efficiencies agree, there is a 
large difference relative to grc4f.
This difference stems mainly from muons with momentum below 3~GeV, which 
have a small detection efficiency, and of which there are more in EXCALIBUR
and KORALW than in grc4f.
As EXCALIBUR and KORALW provide very similar efficiencies, the 
difference between grc4f and the mean efficiency of EXCALIBUR and KORALW
is averaged over all centre-of-mass energies which leads to  a
systematic error of 7.8\%.

\subsection{Background}
\label{sec:background}

The same systematic studies as for the signal Monte Carlo
are performed for the background Monte Carlo. 
The systematic uncertainties derived are given
in Table~\ref{tab:syst}.
In addition for the $\qqee$ analysis
for the multi-hadronic events, a comparison of KK2f and PYTHIA
resulted in a 4.5\% difference in the number of selected events, 
which is also included  in the systematic error.
Due to the small amount of background in the final sample,
these uncertainties 
result in only  a 1.1\% systematic error on the cross-section.

For the process $\ee\to\qqee$, the cross-section of the background 
stemming from
multi-peripheral processes is not well known and
the final states for background and signal
are identical. But the angular distributions, especially of the electrons, are
quite different. 
The generator TWOGEN gives a good relative description
of the electron angular distribution after applying cuts 4, 5 and 6.
From the generator PHOJET, no events are expected after cut 4.
This is not in good agreement with the data.
Therefore, half of the background predicted by TWOGEN is used 
in the calculation of the cross-section and in Tables~\ref{tab:qqee1},
\ref{tab:qqmm1} and \ref{tab:final}.
A systematic uncertainty of 100\% is assigned to this background,
covering both the full TWOGEN and the PHOJET 
prediction. This leads to a 10.0\% systematic error on the measured 
cross-section.

For the $\qqmm$ selection, there is very little background in total.
An error on the
multi-hadronic background of 9.6\% is assigned from the comparison of 
KK2f and PYTHIA. All of the systematic errors taken together result in a 
systematic error on the cross-section of 0.9\%.

\section{Results and Discussion}

With the event selections described in Section~\ref{sec:selection},
the number of data and expected background events are determined 
at several centre-of-mass energies, as given in Table~\ref{tab:final}.
The signal efficiency is calculated by applying the signal selection to
the signal Monte Carlo samples. For  the process $\ee\to\qqee$,
the efficiency is slightly higher for the signal produced with
$\alpha = 1/137$ than for $\alpha = 1/128$, as the $t$-channel contribution
is smaller in the former.
To account for the difference, the average of the two efficiencies
is used in the calculation of the cross-section.
The efficiencies at each centre-of-mass energy are listed in 
Table~\ref{tab:final}.

\begin{table}[htp]
\begin{center}
\begin{tabular}{|l|r|r|c|c||c|c|} \hline
$\rm \sqrt{s}$ (GeV) & $\int{\cal L}$ d$t$& Data & Background & Efficiency ($\%$) & $ \rm \sigma_{Data}$ (fb) & $ \rm \sigma_{grc4f}$ (fb) \\
& (pb$^{-1}$) & & & & & \\
\hline
\hline
\multicolumn{7}{|l|}{$\ee \to \qqee$} \\
\hline
182.7 & 54.7 & \pz 3 & 
0.6  $\pm$ 0.4 & 37.0 $\pm$ 4.3& 
120 $\pm$ 87 $\pm$ 18 &
166   \\
\hline
188.6 & 174.7 & 21  & 
2.5 $\pm$ 1.7 & 39.9 $\pm$ 4.6& 
265 $\pm$ 67 $\pm$ 41 & 
177   \\
\hline
194.9 & 100.0 & 13 & 
1.4 $\pm$ 0.9 & 41.2 $\pm$ 4.8& 
282 $\pm$ 88 $\pm$ 43 &
180  \\ 
\hline
200.7 & 110.3 & 9 & 
1.5 $\pm$ 1.0& 42.0 $\pm$ 4.9& 
163 $\pm$ 66 $\pm$ 25 &
180   \\ 
\hline
206.1 & 214.5 & 12 & 
1.5 $\pm$ 1.3& 40.6 $\pm$ 4.7& 
121 $\pm$ 41 $\pm$ 19 &
177   \\ 
\hline
\hline
\multicolumn{7}{|l|}{$\ee \to \qqmm$} \\
\hline
182.7 & 54.7 & 6 & 
0.3 $\pm$ 0.1& 41.8 $\pm$ 3.8& 
249 $\pm$ 108 $\pm$ 20 & 163   \\
\hline
188.6 & 174.7 & 13 & 
0.6 $\pm$ 0.1 & 40.9 $\pm$ 3.7& 
175 $\pm$ 51  $\pm$ 14 & 168  \\
\hline
194.9 & 100.0 & 9 & 
0.5 $\pm$ 0.1 & 48.3 $\pm$ 4.3 & 
175 $\pm$ 62 $\pm$ 14 & 168  \\
\hline
200.7 & 110.3 & 11 & 
0.4 $\pm$ 0.1 & 52.0  $\pm$ 4.5& 
184 $\pm$ 58 $\pm$ 15 & 165 \\
\hline
206.1 & 214.5 & 13 & 
1.1 $\pm$ 0.2  & 49.5 $\pm$ 4.3& 
112 $\pm$ 35 $ \pm$ 9 &
160   \\
\hline
\end{tabular}
\end{center}
\caption{\it 
The number of events selected at each centre-of-mass energy
between 183~GeV and 209~GeV for the $\qqee$ and $\qqmm$ channels. 
Also listed are the integrated luminosity $\int{\cal L}$dt, 
the background expectation, the 
selection efficiencies and the measured and expected
cross-section for the processes $ \ee \rightarrow \qqee$ and
$ \ee \rightarrow \qqmm $ within the  signal definition. 
$\alpha = 1/128$ is used in the calculation of the signal cross-section.
For the background and efficiencies the error given is the quadratic sum
of the statistical and systematic error.
For the measured cross-section 
the first error is statistical and the second systematic.
The statistical error on the theoretical expectation of the cross-section is 
less than 1 {\mbox{\rm fb}}.
}
\label{tab:final}
\end{table}

The measured cross-sections at each centre-of-mass energy for the
processes $\ee\to\qqee$  and $\ee\to\qqmm$ within the kinematic region
described in Section~\ref{sec:intro} are given in Table~\ref{tab:final}.
For the process $\ee\to\qqmm$, several generators are used to
calculate the cross-section, and all generators agree within 1\%.

\begin{figure}
\centerline{\epsfig{file=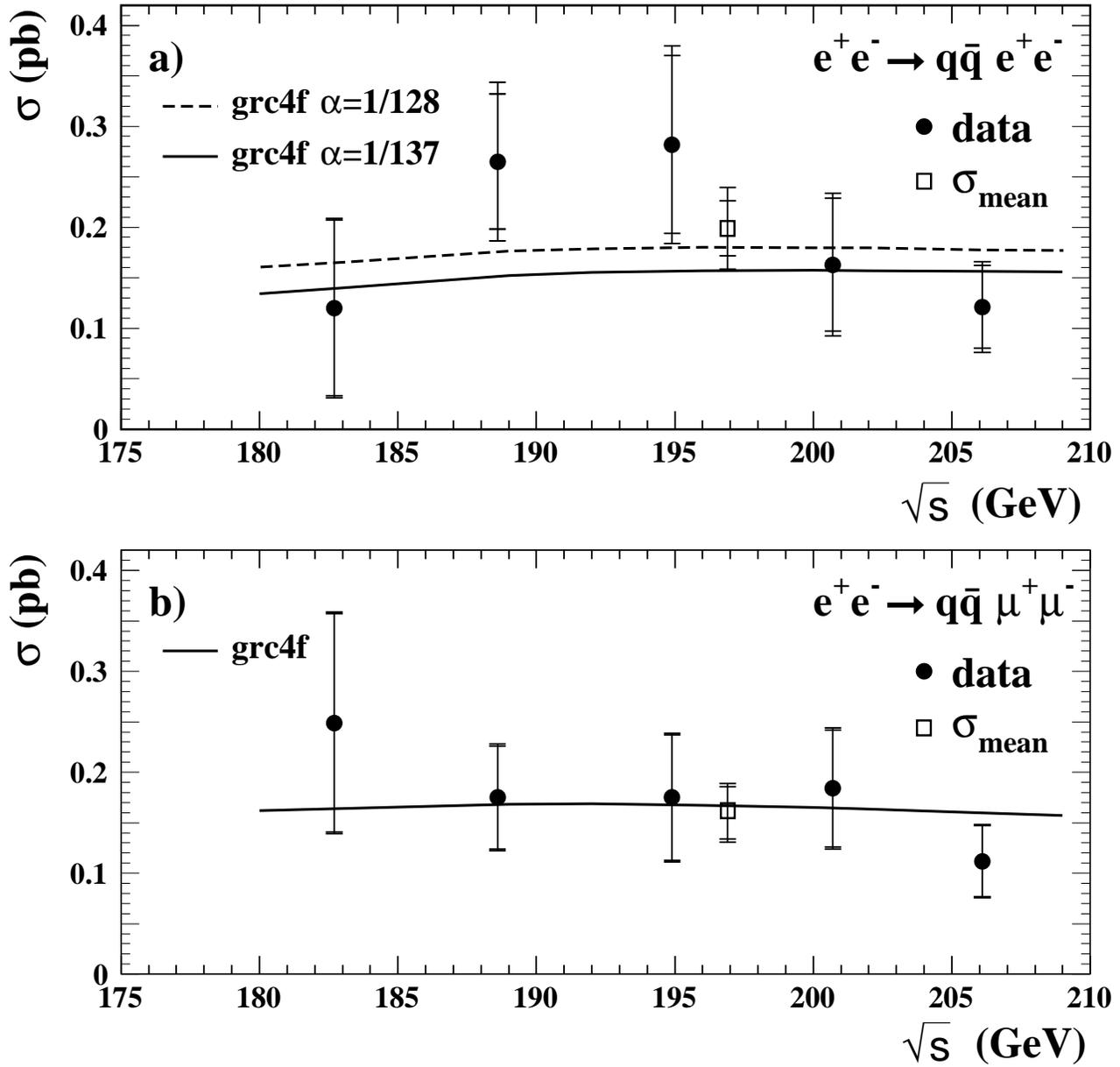,width=1.0\textwidth}}
\caption{ The cross-section for the processes 
(a) $\ee \rightarrow \qqee$  and 
(b) $\ee \rightarrow \qqmm$ 
for the defined signal region.
The dots represent the  measured cross-section at each 
centre-of-mass energy, and the lines give
the prediction from grc4f. 
The open square is the mean cross-section  at $\sqrt{s} = 196.9$~{\rm GeV}. 
The inner error bars represent the statistical error and the outer bars the 
sum of systematic and  statistical error added in quadrature.
}
\label{fig:plots7}
\end{figure}

The measured cross-sections, together with the theoretical predictions,
are shown in Figures~\ref{fig:plots7}a) and b).
The expected values appear to be at their maximum values within the 
studied region, and show
only a small variation with the centre-of-mass energy.
This is in contrast to the cross-section for pair-production of W and Z-bosons,
which shows a very steep rise near threshold.
As the change in the predicted cross-section over the measured range is
much smaller than the error of each individual measurement,
an average cross-section over the whole centre-of-mass energy range
has also been calculated. This average does not take into account the predicted
change with the centre-of-mass energy.
Also, the average cross-section is calculated using 
the expected error at each energy rather than the observed one.
This method gives more reliable results for measurements with a small number
of expected  events.
The  average cross-sections are
$\sigma(\ee \to \qqee)$ = (199 $\pm$ 27 $\pm$ 30)~fb and 
$\sigma(\ee \to \qqmm)$ = (160 $\pm$ 26 $\pm$ 13)~fb.

\begin{figure}
\centerline{\epsfig{file=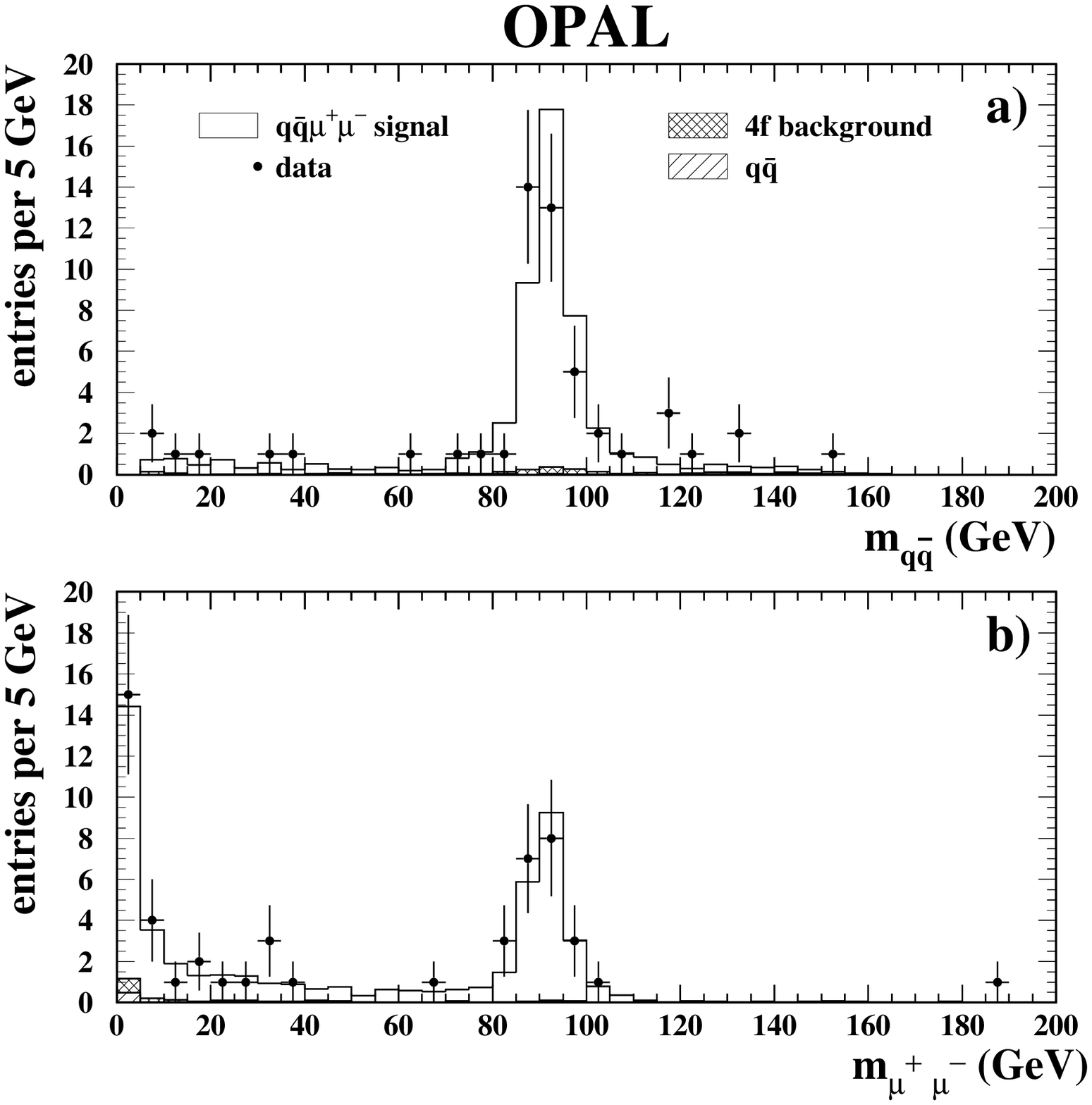,width=1.0\textwidth}}
\caption{ Distribution in the 
$\mqq$ and $\mmumu$ invariant masses obtained from the kinematic fit 
after cut 5 for 
data (dots), signal (open histogram) and background
between 183~GeV and 209~GeV.
The contributions from 
multi-hadronic {\rm ($\qq $)} and four-fermion {\rm (4f)} background events
are shown separately.
}
\label{fig:plots2}
\end{figure}

For the process $\ee\to\qqmm$, only $s$-channel diagrams contribute, and the 
final states are produced via $\ee \to \zz, \zg, \gsgs$. 
This can be seen in the distribution of the invariant masses $\mqq$ and 
$\mmumu$ in Figures~\ref{fig:plots2}~a) and b).
The distribution of $\mmumu$ shows two peaks, one at zero, stemming from the
$\gamma^*$, and one at $\mz$ from the Z decays.
The data are well described by the MC. For the distribution of $\mqq$
the peak around the Z mass is dominant, as the branching ratio of the Z into
quarks is much larger than that into charged leptons.
One event is observed at a very large mass $\mmumu = 188$~GeV. The probability 
to observe at least one event above an invariant mass of 110~GeV is 48\%.

\begin{figure}
\centerline{\epsfig{file=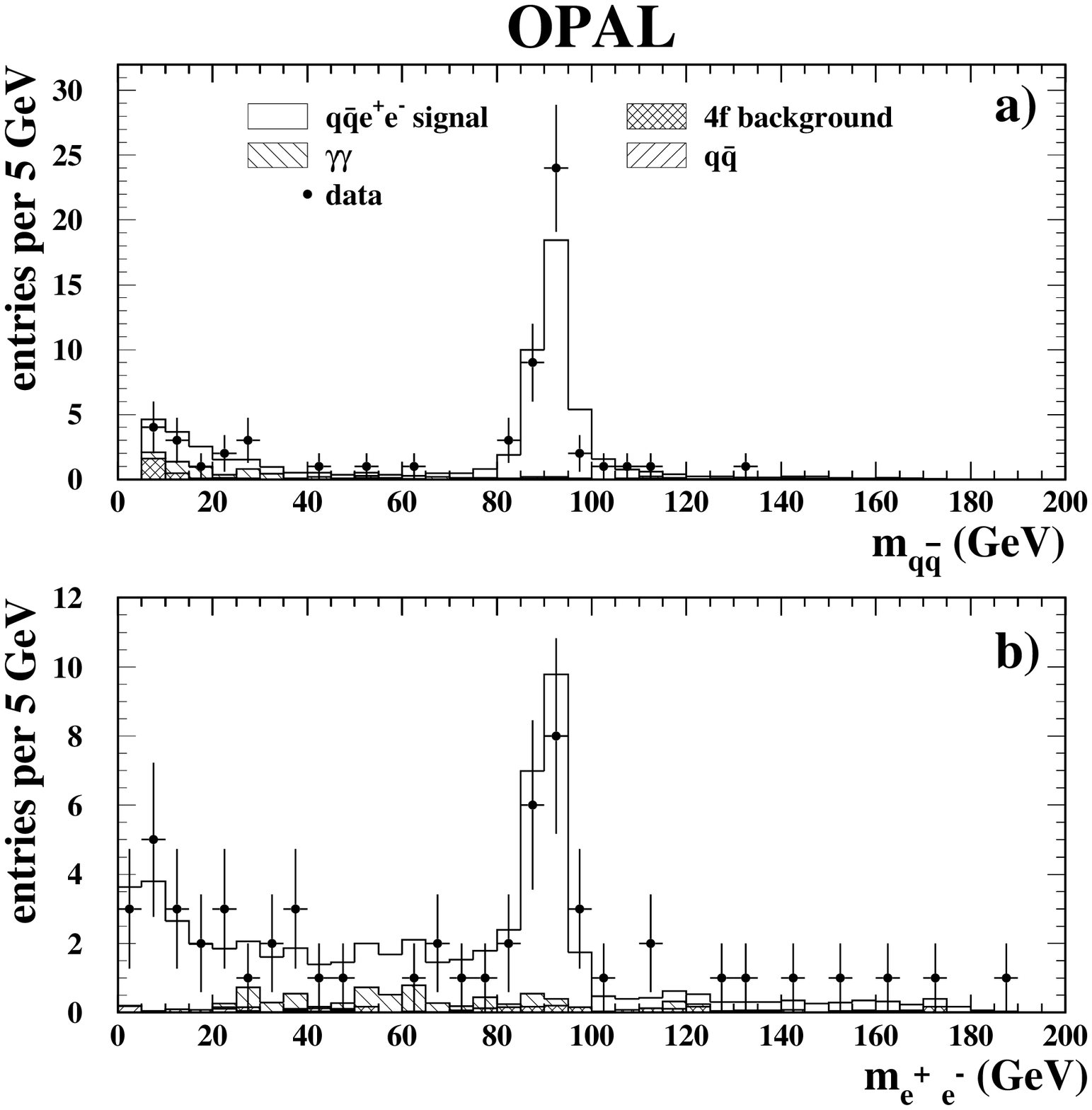,width=1.0\textwidth}}
\caption{ Distribution in the 
$\mqq$ and $\mee$ invariant masses obtained from the kinematic fit 
after cut 6 for data (dots), signal (open histogram) and background
between 183~GeV and 209~GeV.
The contributions from
multi-peripheral {\rm ($\gamma \gamma$)}, multi-hadronic {\rm ($\qq $)}
and four-fermion {\rm (4f)} background events are shown separately.
}
\label{fig:eplot3}
\end{figure}

In the process $\ee\to\qqee$,  $t$-channel diagrams contribute in addition 
to the $s$-channel diagrams. 
The distribution of the invariant masses $\mqq$ and 
$\mee$ are shown in Figures~\ref{fig:eplot3}~a) and b).
The $t$-channel contribution is clear in the distribution of $\mee$.
In contrast to $\mmumu$, there are several events observed at invariant masses
well above the Z mass.

\section{Conclusions}

The cross-sections for the processes $\ee\to\qqee$  and $\ee\to\qqmm$
have been measured 
at centre-of-mass energies between 183~GeV and 209~GeV,
the highest centre-of-mass energies studied to date.
Within chosen kinematic limits,
the predictions from grc4f are in good agreement with the measurements.
The distributions of the invariant masses of the fermion-antifermion pairs
show the expected behaviour, with the $t$-channel contribution clearly visible 
in the $\qqee$ channel.


\medskip
\bigskip\bigskip\bigskip
\appendix
\par
\section*{Acknowledgements}

\par
We particularly wish to thank the SL Division for the efficient operation
of the LEP accelerator at all energies
 and for their close cooperation with
our experimental group.  In addition to the support staff at our own
institutions we are pleased to acknowledge the  \\
Department of Energy, USA, \\
National Science Foundation, USA, \\
Particle Physics and Astronomy Research Council, UK, \\
Natural Sciences and Engineering Research Council, Canada, \\
Israel Science Foundation, administered by the Israel
Academy of Science and Humanities, \\
Benoziyo Center for High Energy Physics,\\
Japanese Ministry of Education, Culture, Sports, Science and
Technology (MEXT) and a grant under the MEXT International
Science Research Program,\\
Japanese Society for the Promotion of Science (JSPS),\\
German Israeli Bi-national Science Foundation (GIF), \\
Bundesministerium f\"ur Bildung und Forschung, Germany, \\
National Research Council of Canada, \\
Hungarian Foundation for Scientific Research, OTKA T-029328, 
and T-038240,\\
Fund for Scientific Research, Flanders, F.W.O.-Vlaanderen, Belgium.\\




\clearpage


\clearpage

\clearpage

\clearpage

\end{document}